# A Novel Decision Tree for Depression Recognition in Speech

Zhenyu Liu[1], Dongyu Wang[1], Lan Zhang[4] and Bin Hu* [1,2,3]

1. Gansu Provincial Key Laboratory of Wearable Computing, School of Information Science and Engineering, Lanzhou University, Lanzhou, China
2. CAS Center for Excellence in Brain Science and Intelligence Technology, Shanghai Institutes for Biological Sciences, Chinese Academy of Sciences, Shanghai, China
3. Joint Research Center for Cognitive Neurosensor Technology of Lanzhou University & Institute of Semiconductors, Chinese Academy of Sciences, Beijing, China
4. Lanzhou University Second Hospital, Lanzhou, China

liuzhenyu@lzu.edu.cn, wangdy18@lzu.edu.cn, zhl2526@126.com, bh@lzu.edu.cn

**Abstract**—Depression is a common mental disorder worldwide which causes a range of serious outcomes. The diagnosis of depression relies on patient-reported scales and psychiatrist's interview which may lead to subjective bias. In recent years, more and more researchers are devoted to depression recognition in speech , which may be an effective and objective indicator. This study proposes a new speech segment fusion method based on decision tree to improve the depression recognition accuracy and conducts a validation on a sample of 52 subjects (23 depressed patients and 29 healthy controls). The recognition accuracy are 75.8% and 68.5% for male and female respectively on gender-dependent models. It can be concluded from the data that the proposed decision tree model can improve the depression classification performance.

## I. INTRODUCTION

Depression is a kind of affective disorder, and its main clinical manifestations are persistently significant negative emotions. The World Health Organization (WHO) listed depression as one of the most serious causes of disability in the world. According to the prediction of the World Health Organization, depression will become the second leading cause of disability in the world by 2030 [1]. Worse still, the cost of treatment for depression is expensive. According to Olesen et al., in 2010, about 30.3 million people were affected by depression, with an average of 797 euros per person in the treatment of depression and 464 euros in indirect medical expenses in the EU [2]. A common diagnostic method is to use depression scales ,like Patient Health Questionaire (PHQ-9)[3], and Hamilton Depression Scale (HAM-D)[4]. These scales are based on the honesty of patients, so there is a certain risk for data authenticity . In order to solve this

problem，scientists have developed some more objective indicatorssuch as：eye movement [5], facial expression [6],and voice [7]. Compared with others, detecting depression through voice signal has advantages of convenience, cheapness, and non-invasiveness.

Previously, some studies have explored some speech-based features that can distinguish between healthy people and patients. In 2015, Yasin ozkanca et al. [8] selected speech production and prosody features , with using neural networks, to reduce the error by 25% in terms of mean absolute error (MAE) and root mean square error (RMSE) on AVEC-2014 date set. . In 2017, Gábor Kiss and Klára Vicsi [9] found that prosody-related features can effectively identify depression. The accuracy of reading speech and interview speech reached 83% and 86%，respectively. In 2018, Meysam Asgariden et al. [10] constructed a new algorithm for extracting speech features. In the task of detecting clinical depression, the accuracy of the features compared to OpenSMILE increased by 9%．In 2018, Aditi mendiratta [11]extracted the MFCC features from normal and depressed subjects' speech respectively, and used self-organization mapping (SOM) algorithm for clustering, with the accuracy of 80.67%. In 2012 Kuan Ee Brian Ooi et al. [12] constructed an early method for predicting depression in adolescents, and studied the individual effects of four speech features on identifying depression. The results show ed that using only glottic or prosodic features, the accuracy is higher than the average. The multi-channel method using four features achieves an accuracy of 73%. More and more studies use speech signals to identify depression.

Some studies have tried to improve model performance by trying different ways. In 2017, Gábor Kiss [13] proposed a multi-speech-based depression detection method using three languages: German, Hungarian and Italian. The results showed that an automatic multiple languages diagnostic tool can be established to detect depression. In 2018, Lang Hea et al. [14] proposed a method combining manual and deep learning features, and augmenting data to solve small sample problem. The validity and robustness of this method is proved by the experiments performed on AVEC2013 and AVEC2014 databases. In 2017, Zhao Jian [15] proposed a multi modal fusion algorithm based on speech signals and facial image sequences for the diagnosis of depression. In 2017, Jiang Dong mei [16] also proposed a multimodal fusion framework composed of deep convolutional neural network (DCNN) and deep neural network (DNN) models, which consider audio, video and text together. In 2018, F.Scibelli, et al[17] proposed a method for automatically identifying depression based on non-text speech features, and the results showed that the classification accuracy rate was greater than 75%. It can be seen that the universality and accuracy of recognition model have been improved through different fusion methods.

We can conclude from the previous studies that the accuracy of model can be improved by variety fusion ways. Therefore, our research focuses on finding a multi-segment fusion method with strong generalization ability and high performance to identify depression.

We will also compare it with the multi-segment fusion methods previously studied. In order to explain which paradigm fusion methods can mostly improve universality and accuracy, we will calculate the accuracy of each paradigm separately, at last we state our conclusions.

The structure of the rest of this paper is as follows: the second section mainly introduces database and data preprocessing, the third section mainly introduces extraction features and classification, the fourth section shows the experimental results, the fifth section is the discussion part, and the sixth section gives the conclusion.

## II. SPEECH DATABASE

On the basis of careful study of previous studies, we designed a data acquisition scheme. We recruit 52 Chinese people (23 depressed patients and 29 healthy controls) –to participate in our experiment. The data contained in this publicly available data set are collected at the Second Affiliated Hospital of Lanzhou University. This data set has the following characteristics:

1. Each participant has , a total of 29 recordings, which are under three different emotional stimulus (positive, neutral and negative).

2. The healthy and depressed subjects were basically matched by gender, age and education level. And, the amount of subjects in each group are balanced.

3. All speech data are collected by using high-quality equipments.

The data set of this disclosure includes two parts: the healthy group and the depressed group, including 29 cases in the healthy group and 23 cases in the depressed group, aged from 18 to 55.（See TABLE I）. All patients were in accordance with DSM-IV [18], and before admission, they were diagnosed by mini [19], and required to speak fluent Chinese and sign informed consent.

TABLE I. Demographic data of male and female subjects

|  | Healthy controls | Depression patients |
|---|---|---|
| Male | 20 | 16 |
| Female | 9 | 7 |

Four types of voice data are collected in the experiment, namely interview voice, passage reading, vocabulary reading and description pictures. There are 29 questions in total. The details are as follows:

Interview: there are 18 questions from several commonly used depression assessment scales, such as DSM-IV [20] and Hamilton Depression Scale [4]. , there are six positive questions, six neutral questions and six negative questions. During the whole recording process, the speaker answers the questions independently according to the specified questions without any intervention from others.

Passage reading: it contains an allegory story "north wind and sun" [21], and the speaker reads the full text according to the prompts on the screen.

Vocabulary reading: it includes six groups of words, including two groups of positive materials, two groups of neutral materials and two groups of negative materials. Each group contains ten words. The speaker reads the materials according to the usual speed of speech according to the prompts.

Picture Description: it contains four pictures, the first three pictures are from the Chinese Facial Affective Picture System [22], which are used as positive, neutral and negative stimulus materials, and the last picture is from the Thematic Apperception Test. The speaker describes the picture content in turn and guesses why the characters in this picture are in this state.

The experimental acquisition site is provided by the Second Affiliated Hospital of Lanzhou University. Before the experiment, the noise in the test room is under 60dB. The voice acquisition software is Adobe Audition CS6, the microphone used is Newsmy TLM102, the sound card is RME FIREFACEUCX, the recording sampling frequency is 44.1 KHz, the sampling depth is 24bit, and the mono channel. The experimental flow is shown in Fig 1.

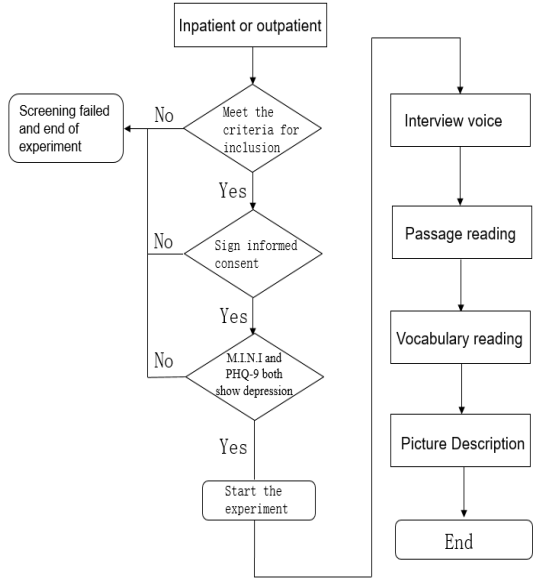

Fig. 1. Experimental flowchart

### III. METHODOLOGY

#### A. Data preprocessing and feature extraction

Before processing the speech data, we performed data preprocessing first, including pre-emphasis, windowing and framing. Pre-emphasis is to enhance the high-frequency component of the signal at the beginning of the transmission line to compensate for the excessive attenuation of the high-frequency component during transmission. The pre-emphasis formula is as follows:

$$H(z) = 1 - \mu z^{-1} \quad (1)$$

μ in this article is taken as 0.97.

The speech signal is a time-varying signal, but we consider the speech signal in a short time as a stable, time-invariant signal. This short time is called the frame, which usually ranges from 10 to 30 ms. This article defines a frame as 25 ms. Framing is achieved by using a weighted method of a movable finite-length window. The window function we use is the Hamming window. Multiply the window function w(n) with the signal s(n) to get the windowed speech signal:

$$s_w(n) = s(n) * w(n) \quad (2)$$

Hamming window of length N:

$$w(n) = \begin{cases} 0.54 - 0.46\cos[2\pi n/(N-1)], & 0 \leq n \leq (N-1) \\ 0, & n = 其他 \end{cases} \quad (3)$$

In this paper, 1609 dimensional features are extracted, among which 16 dimensional features are extracted from the silence segment [23], and the features extracted from the acoustic segment are divided into three parts. The tremor features are extracted manually, totaling 8 dimensions. Through Open SMILE [24], 1582 dimensional features and three energy related features are extracted: energy, low frequency ratio and temporal time. The short-time energy can represent the change of speech signal energy with time. It reflects the change of speech signal amplitude, which is helpful to recognize depression. Each feature is explained as follows:

The features that are extracted from the silent segment are also called pause features, or each pause reflects the brain response of the speaker at that time. To a certain extent, pause features can distinguish normal and depression people [23]. The silent part of the speech segment is extracted by speech endpoint detection [25], as shown in Fig 2, and the pause feature [26] is introduced as shown in TABLE II.

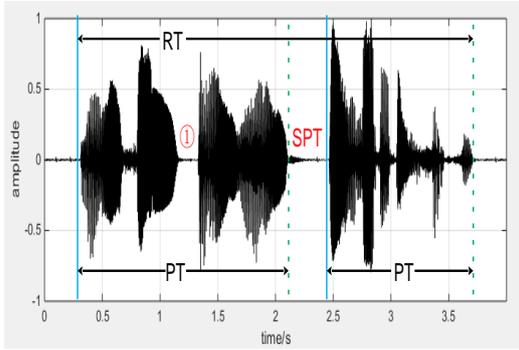

Fig. 2. speech endpoint detection method extracts the features of the silent part. The beginning of speech is represented by solid line, and the end of speech is represented by dotted line. SPT stands for pause. Because ① is too short, it is directly discarded.

TABLE II. Introduction of pause characteristics

| Feature Name | Meaning |
| --- | --- |
| Max Duration | Maximum pause time |
| Reaction Time | reaction time |
| Total Recording Time | Recording duration |
| Total Vocalization Time | Duration of sound part |
| Total Pause Time | Duration of silent part |
| Number Of Pause | Pause times |
| Mean Pause Length | Average duration of pause |
| Percent Pause Time | Proportion of silent parts |
| Speech Pause Ratio | Proportion of sound part |
| Total Recording Time | Total sequence length |
| Total Vocalization Time | Continuous sound duration |
| Total Pause Time | Continuous silent period |
| Number of Pause | Number of sound and silence changes |
| Mean Pause Length | Average duration of sound section |
| Percent Pause Time | Average duration of silent segment |
| Speech Pause Ratio | The ratio of the length of sound period to the length of silent period |

The features of voice segment are extracted by Open SMILE [24], such as MFCC, LPC, jitter and their correlation statistics. In addition, the tremor feature extracted manually is the correlation statistics calculation of the 15 Hz modulation of the amplitude and fundamental frequency features.

B. Feature selection

In order to get more useful information for identifying depression, we extracted 1609 dimensional features. For improving the generalization ability of the model, we need to reduce the dimension. This paper chooses feature selection algorithm to reduce the dimension. Since it can both maintains the original physical meaning of the feature, and can also remove some "irrelevant features" and "redundant features", strengthen the model training speed, and reduce over fitting.

Relief [28] is a feature weight algorithm, which randomly selects feature Q, finds k-nearest-neighbor sample W from the same set as feature Q, finds k-nearest-neighbor sample e from another set, and finally updates the feature weight. Features that are less than the threshold will be removed. This algorithm selects the features that can represent health and depression as much as possible, which helps us achieve the goal of binary classification. Moreover, the relief algorithm is simple and easy to implement, and it is more efficient than other feature selection algorithms. Therefore, this paper selects the relief algorithm for feature selection.

C. Classifiers

The classifier is trained on the existing data to obtain a classification model for prediction. This paper only selects two common classifiers to evaluate features: support vector machine (SVM) [29] and random forest (RF) [30]

SVM is a binary classification model, which is suitable for classification of small samples. Its classification idea is to seek a hyperplane for classes A and B so that the distance between the points closest to the hyperplane is as large as possible.

Decision tree is a basic and common classifier. It is a tree structure. Each node represents an attribute, each branch finally outputs a test, and each leaf node represents a category.

This paper uses 4 fold cross validation for model evaluation. The 4-fold cross-validation is to divide the data set into 4 parts, and each 3 parts is used as the training set, and then use the Relief algorithm to obtain the features with the highest scores in the first 20 dimensions. We use SVM and random forest classifiers for training, and the remaining part as a test set, features use the features from the training set, perform fifty classifications, and then average the classification results.

D. Fusion Method

The simple voting method is that twenty nine segments are calculated the classification results of the same subject respectively, and we count the times of normal and depression respectively. The results with more times are the final classification results.

We refer to the method of reference [31], which is to construct a binary tree for speech segment fusion through the correct rate, and we take the speech segment with the highest correct rate as the root node. If it is judged normal, we select the speech segment with the highest accuracy and relatively high specificity as the left child in the remaining speech segments, otherwise, we select the speech segment with the highest accuracy and relatively high sensitivity in the remaining speech segments. For the same subject, the binary tree model predicts depression accumulation twice. Stopping building and outputting depression. We predict health in the same way.

In this paper, we propose a binary tree fusion method. The results of previous studies [31] [32] show that for men and women, different behavioral indicators have different contributions to depression and post-traumatic stress disorder, and a fusion method based on decision trees can improve the recognition accuracy of depression.

Therefore, we propose a new tree fusion scheme. It's different from the second method. Our method is an automatic method to fuse SVM models of multiple speech segments, in which the root node is the most sensitive SVM model, and the subsequent nodes are also selected according to the sensitivity and specificity.

We use specificity and sensitivity to build a binary tree for speech segment fusion. The binary tree is a tree structure with a maximum of two sub trees at each node. Usually, the sub trees are called left subtree and right subtree. Sensitivity refers to the probability of not being missed when diagnosing a disease, and specificity refers to the probability that there is no misdiagnosis when diagnosing a disease. Therefore, we use specificity and sensitivity to evaluate the ability to identify health and depression, respectively. We use this binary tree to simulate the process of diagnosing depression in the scale. The modeling process is to set the most sensitive speech segment as the root node. If the judgment is normal, the most specific speech segment is selected as the left child among the remaining speech segments. Otherwise, the most sensitive speech segment is selected as the right child among the remaining speech segments. For the same subject, the binary tree model predicts health cumulatively twice, that is, stops building, outputs healthy results, and predicts depression for the same way. Each node of the binary tree is an SVM model of the selected speech segment, which is used to output the result. A binary tree has a maximum of four layers, so it is fast to model and has a certain generalization ability. The pseudocode for creating a binary tree is as follows:

```
  Begin
 health = 0;// Number of times is that
            the model judged that
            the subject was healthy
 depression = 0; // Number of times
                is that the model
                judged that the
                subject was
                depression
 // Assign the most sensitive speech
segment to the root node
  Node.data = max(sensitivity);
  while(health < 2 || depression <2)
        if(SVM(node.data) == 1)
         // Select the most specific
           speech segment among
           the remaining speech
           segments as the left child
            node.left.data =
               max(specificity);
            node = node.left;
            health++;
        else
        // Select the most sensitive
           speech segment among
           the remaining speech
           segments as the right
           child
           node.right.data=
               max(sensitivity);
            node = node.right;
            depression ++;
  // Output result
  if health == 2
      return health；
  else
      return depression;
  end
 end
```

## IV. EXPERIMENTS AND RESULTS

Previous studies have proved that paradigm fusion can improve the accuracy of the model. Different paradigms have different abilities to recognize health and depression. The accuracy can only represent the overall performance of the paradigm. We need to use specificity and sensitivity to evaluate the ability of the paradigm to recognize health and depression respectively. Therefore, we assume that the performance of speech segment fusion can be improved by building a binary tree with specificity and sensitivity. In this experiment, three fusion methods are used to fuse the classification results of each speech segment. Among them, the feature is the twenty dimensional feature with the highest score selected by relief algorithm.

We calculate the performance of each speech segment separately as a comparison with the results of the fusion method. The process is as follows:

First, we divide the data set into 4 parts by using 4 fold cross validation, each 3 parts are used as the training set, then we get the top 20 dimension features by Relief algorithm, use SVM and random forest classifiers for training, the remaining one is used as the test set, and use the features from the training set for 50 times of classification, and finally average the classification results.

The results are shown in Fig 3, Fig 4，Fig 5 and TABLE III. On the whole, the performance of SVM classifier is similar to that of random forest classification, and SVM is slightly better than random forest classification. Therefore, the SVM classifier is used to obtain the prediction results of each speech segment in the speech segment fusion.

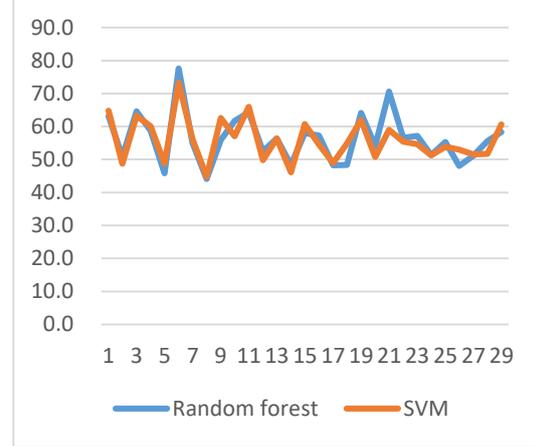

Fig 3. The classification accuracy of all speech segments of all participants (%)

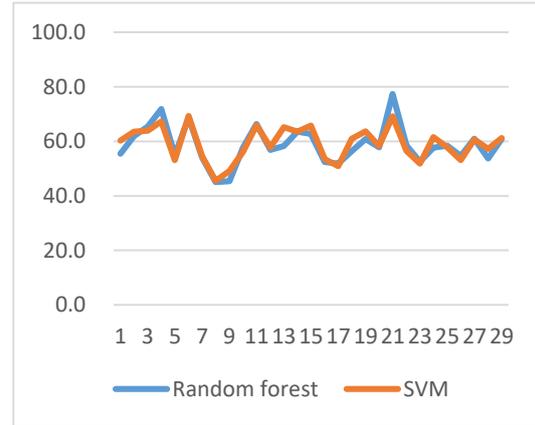

Fig 4. The classification accuracy of all speech segments of male subjects (%)

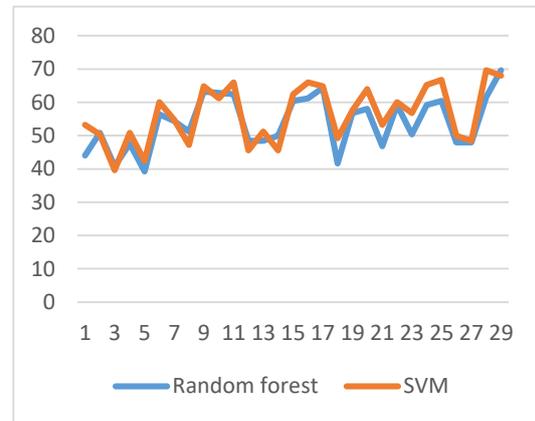

Fig 5. The classification accuracy rate of

all speech segments of female participants

TABLE III. Accuracy of four experimental scenarios (%)

|  | Interview | Passage reading | Vocabulary reading | Picture Description |
|---|---|---|---|---|
| All people | 56.8 | 62.0 | 54.2 | 54.2 |
| Male | 59.2 | 63.6 | 58.4 | 58.0 |
| Female | 54.2 | 57.6 | 61 | 60.1 |
| Average | 56.7 | 61.1 | 57.9 | 57.4 |

We use the three fusion methods described in the third section to fuse the SVM model of 29 speech segments, cycle 50 times, and average the results. The results of the above three methods are shown in TABLE III, TABLE IV and TABLE V. At the same time, we listed the average and maximum values of the accuracy rate of the 29-segment speech, as well as the accuracy for reading short passages. Obviously, the third method is better than the other two fusion methods.

TABLE IV. Comparison of fusion results of speech segments of all subjects

| Method | Accuracy |
|---|---|
| Average accuracy of 29 speech segments | 55.9 |
| The highest accuracy of 29 speech segments | 73.2 |
| Passage reading | 62.0 |
| Voting fusion of 29 voice segments | 57.0 |
| Binary tree fusion model based on accuracy | 66.8 |
| A new design of binary tree fusion model | 70.0 |

TABLE V. Comparison of fusion results of speech segments of male subjects

| Method | Accuracy |
|---|---|
| Average accuracy of 29 speech segments | 59.2 |
| The highest accuracy of 29 speech segments | 69.3 |
| Passage reading | 63.6 |
| Voting fusion of 29 voice segments | 55.3 |
| Binary tree fusion model based on accuracy | 70.5 |
| A new design of binary tree fusion model | 75.8 |

TABLE VI. Comparison of fusion results of speech segments of female subjects

| Method | Accuracy |
|---|---|
| Average accuracy of 29 speech segments | 56.4 |
| The highest accuracy of 29 speech segments | 69.6 |
| Passage reading | 57.6 |
| A new design of binary tree fusion model | 49.1 |
| Binary tree fusion model based on accuracy | 63.2 |
| A new design of binary tree fusion model | 68.5 |

V. DISCUSSIONS

This paper studies the method of speaker (depression / Health) classification by speech, and designs a binary tree fusion model. The results show that this fusion model has better performance than other fusion models, and has stronger generalization ability compared with a single paradigm.

We obtained 20-dimensional features through feature selection, and then calculated the classification performance of 29 speech segments. In the three groups, for a single speech segment, the sixth and eleventh paragraphs in the interview performed better . th possible reason is    the interview is a question that enables the subjects to express their ideas freely,  so that the interview can identify depression better . Although the interview can get the highest accuracy, passage reading shows higher accuracy, as shown in TABLE III. It may be due to the slow response of depression patients. When they are surrounded by unfamiliar equipments, their response time is longer than that of healthy subjects, this can be caught by     the pause feature classification, on the other hand, essay reading contains the most information, which is also good for classification.

The binary tree designed in this paper is fused in the male paradigm, which is higher than the maximum accuracy rate of 29 speech segments, while the other two groups are similar to the maximum accuracy rate of 29 speech segments. This may be due to the small number of female subjects, which reduces the generalization ability of the model. In addition, the three groups are higher than the average of the accuracy rate of 29 speech segments and the accuracy rate of reading paragraphs. Therefore, the newly designed binary tree fusion method in this paper performs well.

At last, we compare different speech segment fusion methods, among which 29 speech segments are the worst, which may be due to too many paradigms of poor classification effect, interfering with the prediction of the results by the model. The binary tree fusion designed in this paper has the best effect, because the essence of the binary classification problem is to divide them into two categories, and the specificity and sensitivity represent the diagnosis rate of healthy people and the diagnosis rate of depressed people respectively, which are the prior probability of health and depression respectively. Therefore, after the introduction of specificity and sensitivity, the new fusion method is more targeted to identify depression than the second method, and the fusion results will be improved. However, because the newly designed binary tree fusion method is based on a priori probability, a large amount of data is needed for training set. In this paper, the number of subjects is small, which may reduce the generalization ability of the model. Overall, the newly designed binary tree fusion method performs better than the other two methods.

## VI. CONCLUSION

In this paper, we used machine learning to build a speech-based predictive depression model on this publicly available speech data set. Through the binary tree fusion model designed in this study, The recognition accuracy are 75.8% and 68.5% for male and female respectively on gender-dependent models. It can be concluded that the proposed decision tree model can improve the depression classification performance. In view of the shortcomings of this study, in the future work, we will collect more voice information of participants to help the research. In addition, we will further optimize the speech-based predictive depression model, improve the accuracy and generalization ability of the model, and look for features that are more helpful for classification.

ACKNOWLEDGMENT

This work was supported in part by the National Key Research and Development Program of China (Grant No. 2019YFA0706200), in part by the National Natural Science Foundation of China (Grant No.61632014, No.61627808, No.61210010, No.61802159), in part by the National Basic Research Program of China (973 Program, Grant No.2014CB744600),and in part by the Program of Beijing Municipal Science & Technology Commission (Grant No.Z171100000117005).